\documentclass[final,5p,times,twocolumn
]{elsarticle}

\usepackage[T1]{fontenc}
\usepackage[utf8]{inputenc}

\usepackage{amsmath}
\usepackage{amsfonts}
\usepackage{amssymb}
\usepackage{amsxtra}
\usepackage{array}
\usepackage{svg}
\usepackage{graphicx}
\usepackage{hepunits}
\usepackage{svg}
\usepackage{color}
\usepackage{tikz}
\usepackage{cuted} 
\usepackage[linkcolor=blue,citecolor=blue,urlcolor=blue,colorlinks=true,breaklinks]{hyperref}
\usepackage{xspace}
\usepackage[normalem]{ulem}

\usepackage[mathscr,scaled=1.15]{urwchancal}
\usepackage{relsize}

\DeclareMathAlphabet{\mathcal}{OMS}{cmsy}{m}{n}

 \def\srm#1{\rm{\scriptstyle{#1}}}

\def\fig#1{Fig.~\ref{#1}}

\def\g{\Gamma}  
\def\s#1{{\scriptscriptstyle #1}}
\def\1eq#1{Eq.~(\ref{#1})}

\def\x{x}
\def\y{y}

\def\Tff{D_{\!\text{\smaller{4g}}}}   
\def\Tffarg{\Tff(s^{2}\!,R,\phi)}    
\def\Tffargi{\Tff(s^{2}\!,R_{i},\phi_{i})}    

\def\Dav{\overline{D}}      
\def\DavL{\overline{D}_{\!\textit{\smaller{L}}}}    
\def\Davn{\overline{D}_{\!\textit{\smaller{N}}}}   

\def\gbar{{\overline \Gamma}\vphantom{\Gamma} }

\newcommand{\Ls}{ \mathit{L}_{{sg}}}

\def\2eqs#1#2{Eqs.~(\ref{#1}) and~(\ref{#2})}

\newcommand{\be}{\begin{equation}}
\newcommand{\bea}{\begin{eqnarray}}
\newcommand{\ee}{\end{equation}}
\newcommand{\eea}{\end{eqnarray}}

\def\1eq#1{Eq.~(\ref{#1})}

\def\2eqs#1#2{Eqs.~(\ref{#1}) and~(\ref{#2})}
\def\3eqs#1#2#3{Eqs.~(\ref{#1}),~(\ref{#2}) and~(\ref{#3})}

\def\ie{{\it i.e.}, }
\def\eg{{\it e.g.}, }

\def\sfg{\bar{s}}

\newcommand{\fatg}{{\rm{I}}\!\Gamma}

\newcounter{comment}

\newcommand{\FScolor}{red}

{\refstepcounter{comment}%
\begin{quote}
\color{\PDcolor}{#1~ \normalsize \textbf{\underline{Comment} $\sharp$\thecomment~by~FP:~}}}%
{\end{quote}}

{\refstepcounter{comment}%
\begin{quote}
\color{\FScolor}{#1~ \normalsize \textbf{\underline{Comment} $\sharp$\thecomment~by~FS:~}}}%
{\end{quote}}

{\refstepcounter{comment}%
\begin{quote}
\color{\CMcolor}{#1~ \normalsize \textbf{\underline{Comment} $\sharp$\thecomment~by~Pp:~}}}%
{\end{quote}}

 \biboptions{sort&compress}

\journal{Physics Letters B}

\begin{document}

\begin{frontmatter}

\title{
Nonperturbative four-gluon vertex in soft kinematics}

\author[Campinas]{A.~C.~Aguilar}

\author[Sevilla]{F. De Soto}

\author[UNanjing,INP]{M.~N.~Ferreira}

\author[UV,GSI]{J. ~Papavassiliou}

\author[Sevilla]{F. Pinto-G\'{o}mez}

\author[Huelva,Saclay]{J. Rodr\'{i}guez-Quintero}

\author[Campinas]{L.~R. Santos}

\address[Campinas]{
University of Campinas - UNICAMP, Institute of Physics Gleb Wataghin, 13083-859 Campinas, S\~ao Paulo, Brazil}

\address[Sevilla]{ Dpto. Sistemas F\'{i}sicos, Qu\'{i}micos y Naturales, Univ. Pablo de Olavide, E-41013 Sevilla, Spain}

\address[UNanjing]{School of Physics, Nanjing University, Nanjing, Jiangsu 210093, China
}

\address[INP]{Institute for Nonperturbative Physics, Nanjing University, Nanjing, Jiangsu 210093, China
}

\address[UV]{Department of Theoretical Physics and IFIC, University of Valencia and CSIC, E-46100, Valencia, Spain}

\address[GSI]{\mbox{ExtreMe Matter Institute EMMI}, GSI,
Planckstrasse 1, 64291, Darmstadt, Germany}

\address[Huelva]{
Department of Integrated Sciences and CEAFMC, University of Huelva, E-21071 Huelva, Spain}

\address[Saclay]{Irfu, CEA, Université de  Paris-Saclay, 91191, Gif-sur-Yvette, France}

%

%
%


\begin{abstract}
We present a nonperturbative study of the form factor 
associated with the projection of 
the full four-gluon vertex on its classical tensor,
for a set of kinematics 
with one vanishing and three arbitrary 
external momenta. The treatment 
is based on the Schwinger-Dyson equation governing this vertex, 
and a large-volume lattice simulation, involving ten thousand gauge field configurations. 
The key hypothesis employed in both approaches 
is the ``planar degeneracy'', which 
classifies diverse configurations by means 
of a single variable, thus enabling their 
meaningful ``averaging''. 
The results of both approaches show 
notable agreement, revealing a considerable 
suppression of the averaged 
form factor in the infrared.
The deviations from the 
exact planar degeneracy are discussed in detail, 
and a supplementary variable is used to 
achieve a more accurate description.
The effective charge defined through 
this special form factor is computed 
within both approaches, 
and the results obtained are 
in excellent agreement. 

\end{abstract}

\begin{keyword}
Quantum Chromodynamics \sep Four-gluon vertex \sep Lattice QCD\sep Schwinger-Dyson Equations


\end{keyword}

\end{frontmatter}




\section{Introduction}
\label{introduction}

In the last decades our 
comprehension 
of Quantum Chromodynamics (QCD)
~\cite{Yang:1954ek,Gross:1973id,Politzer:1973fx,Marciano:1977su}
has 
improved substantially due to 
the ongoing 
exploration of basic  
correlation functions 
of the theory, such as propagators 
and vertices~\cite{Roberts:1994dr,Alkofer:2000wg,Fischer:2006ub,Binosi:2009qm,Cloet:2013jya,Aguilar:2015bud,Eichmann:2016yit,Huber:2018ned,Papavassiliou:2022wrb,
Ferreira:2023fva}. 
This systematic survey is advancing  
thanks to studies 
based on nonperturbative methods 
formulated in the continuum, such as 
Schwinger-Dyson equations (SDEs)~\cite{Roberts:1994dr,Alkofer:2000wg,Fischer:2006ub,Roberts:2007ji,Binosi:2009qm,Cloet:2013jya,Aguilar:2015bud,Eichmann:2016yit,Huber:2018ned,Papavassiliou:2022wrb,Aguilar:2008xm,Boucaud:2008ky,Eichmann:2009qa,Fischer:2008uz,RodriguezQuintero:2010wy,Huber:2012zj,Gao:2021wun,Huber:2016tvc,Huber:2020keu,Aguilar:2021uwa,Ferreira:2023fva}, the functional
renormalization group~\cite{Cyrol:2014kca,Braun:2007bx,Fister:2013bh,Pawlowski:2003hq,Pawlowski:2005xe,Cyrol:2017ewj,Cyrol:2018xeq,Corell:2018yil,Blaizot:2021ikl,Horak:2021pfr,Pawlowski:2022zhh}, or models that incorporate certain key aspects of 
the theory~\cite{Tissier:2010ts,Dudal:2008sp,Mintz:2017qri,Barrios:2020ubx,Pelaez:2021tpq,Barrios:2024ixj}. 
In addition, large-volume   
gauge-fixed lattice 
simulations~\cite{Cucchieri:2006tf,Cucchieri:2007md,Bogolubsky:2007ud,Bogolubsky:2009dc,Cucchieri:2008qm,Cucchieri:2009zt, Oliveira:2009eh,Oliveira:2010xc,Ayala:2012pb,Oliveira:2012eh,Athenodorou:2016oyh,Duarte:2016ieu,
Sternbeck:2017ntv,Boucaud:2017ksi,Boucaud:2018xup,Aguilar:2021okw,Maas:2011se,Aguilar:2021lke} provide invaluable insights into 
the evolution of correlation functions at intermediate and low values of their physical momenta. 
This combined information is essential for the 
veracious computation of physical 
observables~\cite{Cloet:2013jya,Meyer:2015eta,Eichmann:2016yit,Souza:2019ylx,Huber:2023mls,Pawlowski:2022zhh,Eichmann:2020oqt}, 
and the 
scrutiny of the theoretical 
underpinnings of 
non-Abelian gauge theories
~\mbox{\cite{Cornwall:1981zr,Fischer:2003rp,Cui:2019dwv,Binosi:2014aea,Aguilar:2022thg,Boucaud:2011ug,Aguilar:2010cn,Kondo:2014sta,Greensite:2003bk}}.

 Whereas the two- and three-point sectors 
of QCD have been the focal point of
intense investigation, 
the nonperturbative 
features of the four-gluon vertex, 
$\fatg^{abcd}_{\mu\nu\rho\sigma}(q,r,p,t)$, 
remain largely unexplored;
for perturbative results, see~\cite{Pascual:1980yu,Brandt:1985zz, Papavassiliou:1992ia, Hashimoto:1994ct, Gracey:2014ola, Gracey:2017yfi,Ahmadiniaz:2013rla,Ahmadiniaz:2016qwn}. The main obstacle in the continuum is the proliferation 
of Lorentz and color structures,
while on the lattice 
the statistical noise increases 
considerably as one advances from three to four gluon legs.  
As a result, both SDE studies~\cite{Binosi:2014kka,Huber:2020keu,Aguilar:2024fen} 
and lattice simulations~\cite{Colaco:2024gmt} have been restricted 
to simple 
kinematic setups, where the logistic complexity is vastly reduced; such are the {\it ``collinear''} 
configurations, 
where all momenta are parallel.

In the present work we carry out 
a comprehensive study of the 
four-gluon vertex for a considerably wider set 
of kinematics. 
In particular, 
we consider the 
case where one momentum vanishes ($t=0$),
while the space-like momenta
$q$, $r$, and $p$  are 
arbitrary; 
we will refer to 
these configurations 
as {\it ``soft kinematics''}. 

Our analysis 
is based on the synergy between two distinct 
nonperturbative approaches:
the SDE 
governing the 
evolution of this vertex, and 
gauge-fixed simulations 
performed on  
large-volume lattices.
In both cases, the computations are  
carried out in the {\it Landau gauge}.  

The central theme of our 
considerations is the 
property of \mbox{{\it ``planar degeneracy''}}~\cite{Eichmann:2014xya}, 
which has been extensively studied in the context of the three-gluon vertex 
~\cite{Pinto-Gomez:2022brg,Aguilar:2023qqd,Pinto-Gomez:2024mrk}. In the case of the 
four gluon vertex, this property affirms 
that the form factor
associated with the classical tensor is approximately equal for all configurations lying on the plane 
$\sfg^{2} =  (q^2+r^2+p^2+t^2)/2$, or, 
in the soft kinematics,
$s^2 = (q^2 + r^2 + p^2)/2 $.  
Even though not exact, this 
feature is particularly useful
on the lattice, because configurations 
with the same $s^2$ are treated as equivalent; 
thus, seemingly unrelated measurements are summed up and averaged, leading to a 
vast improvement of the signal. 

The averaged form factor 
extracted from the lattice
displays a clear    
infrared suppression with respect to its
tree-level value (unity), in qualitative agreement with previous continuum studies performed in other kinematic configurations~\cite{Binosi:2014kka,Huber:2020keu,Aguilar:2024fen,Barrios:2024ixj}. Moreover, 
it is in very good agreement 
with the corresponding 
result obtained from a detailed SDE analysis in the soft kinematics, 
where the assumption 
of the planar degeneracy 
has been employed in order 
to simplify the iterative 
procedure.

The deviation of the result from the exact 
planar degeneracy is 
quantified in terms of an 
additional kinematic parameter, which, in conjunction with $s^2$, 
allows for a more accurate 
description of the underlying dynamics. 

Finally, the 
renormalization-group invariant (RGI)
effective charge corresponding to this interaction is constructed, 
using the lattice and the SDE results; the two curves so obtained show excellent agreement.

\section{General structure and kinematics}
\label{sec:tensorial}

The correlation function 
composed out of four gauge fields at momenta $q$, $r$, $p$, and $t$ (with $q + r + p + t = 0$), is defined as 
\begin{eqnarray}
\mathcal{G}_{\mu\nu\rho\sigma}^{abcd}(q,r,p,t) = \langle \widetilde{A}_\mu^a(q)\widetilde{A}_\nu^b(r)\widetilde{A}_\rho^c(p) \widetilde{A}_\sigma^d(t) \rangle \;,
\end{eqnarray}
where $\widetilde{A}_\mu^a(q)$ are the SU(3) gauge fields in Fourier space, and the average  $\langle . \rangle$ indicates functional integration over the gauge space (Monte Carlo average in lattice QCD). 
The function $\mathcal{G}_{\mu\nu\rho\sigma}^{abcd}(q,r,p,t)$ contains 
a connected part, 
denoted 
$\mathcal{\widetilde C}^{abcd}_{\mu\nu\rho\sigma}$, 
and disconnected 
propagator-like
contributions. {Additionally, the amputated vertex $\mathcal{C}^{abcd}_{\mu\nu\rho\sigma}$ can be further separated into one-particle irreducible (1PI) and one-particle reducible (1PR) parts, denoted respectively $\fatg^{abcd}_{\mu\nu\rho\sigma}$ and $\mathbb V^{abcd}_{\mu\nu\rho\sigma}$, as shown in Fig.~\ref{fig:1P1}. At tree-level, 
\mbox{$\fatg^{abcd}_{\mu\nu\rho\sigma}\to  \Gamma^{abcd}_{0\,\mu\nu\rho\sigma}$}, see, \eg 
Eq.~(2.11) in~\cite{Aguilar:2024fen}.

In the Landau gauge 
that we employ throughout, 
the gluon propagator, $\Delta_{\mu\nu}^{ab}(q)=-i\delta^{ab}\Delta_{\mu\nu}(q)$, is given by  
\begin{align}
\Delta_{\mu\nu}(q) = \Delta(q^2) P_{\mu\nu}(q) \,, \qquad 
P_{\mu\nu}(q) = \delta_{\mu\nu} - q_{\mu} q_{\nu}/q^2 \,.
\label{eq:prop}
\end{align}
Then, the amputation of the external legs 
proceeds by setting 
\be
\mathcal{\widetilde C}^{abcd}_{\mu\nu\rho\sigma}(q,r,p,t)
= \Delta(q) \Delta(r) \Delta(p)\Delta(t)\,\overline{\mathcal{C}}^{abcd}_{\mu\nu\rho\sigma}(q,r,p,t) \,,
\label{eq:Ctdefproj}
\ee
where 
\begin{align}
\mathcal{\overline C}^{abcd}_{\mu\nu\rho\sigma}(q,r,p,t) := 
 {\cal T}^{\mu'\nu'\rho'\sigma'}_{\mu\nu\rho\sigma}(q,r,p,t) \,\mathcal{C}^{abcd}_{\mu'\nu'\rho'\sigma'}(q,r,p,t)\,, 
\label{eq:Cbar}
\end{align}
with 
\begin{align}
{\cal T}^{\mu'\nu'\rho'\sigma'}_{\mu\nu\rho\sigma}
(q,r,p,t):= P^{\mu'}_{\mu}(q) P^{\nu'}_{\nu}(r)P^{\rho'}_{\rho}(p)P^{\sigma'}_{\sigma}(t) \,,
\label{projP}
\end{align}
or, equivalently, 
\begin{align}
\mathcal{\overline C}^{abcd}_{\mu\nu\rho\sigma}(q,r,p,t) = & 
-ig^2\overline{\fatg}^{abcd}_{\mu\nu\rho\sigma}(q,r,p,t) + \overline{\mathbb V}^{abcd}_{\mu\nu\rho\sigma}(q,r,p,t) \,,
\label{eq:allbars}
\end{align}
with
\begin{align}
\overline{\mathbb V}^{abcd}_{\mu\nu\rho\sigma}(q,r,p,t) = -i\,\overline\fatg^{ade}_{\mu\sigma\lambda} \,
\Delta^{\lambda\beta} \,
\overline\fatg^{bce}_{\nu\rho\beta} + \mathrm{crossed}\,,
\label{eq:allbars1}
\end{align}
where 
\begin{align}
\overline{\fatg}^{abcd}_{\mu\nu\rho\sigma}(q,r,p,t) =  {\cal T}^{\mu'\nu'\rho'\sigma'}_{\mu\nu\rho\sigma}(q,r,p,t)\fatg^{abcd}_{\mu'\nu'\rho'\sigma'}(q,r,p,t)\,, 
\label{4g_transv}
\end{align}
is the 
\emph{transversely projected} 1PI 
four-gluon vertex, while
\begin{align}
\overline\fatg^{abc}_{\alpha\beta\gamma}(q,r,p)
= P^{\alpha'}_\alpha(q)P^{\beta'}_\beta(r)P^{\gamma'}_\gamma(p)
\fatg^{abc}_{\alpha'\beta'\gamma'}(q,r,p)
\,,
\end{align}
is the transversely-projected three-gluon vertex. 

\begin{figure}[t]
    \centering
\hspace{-0.8cm}\includegraphics[width=1.05\linewidth]{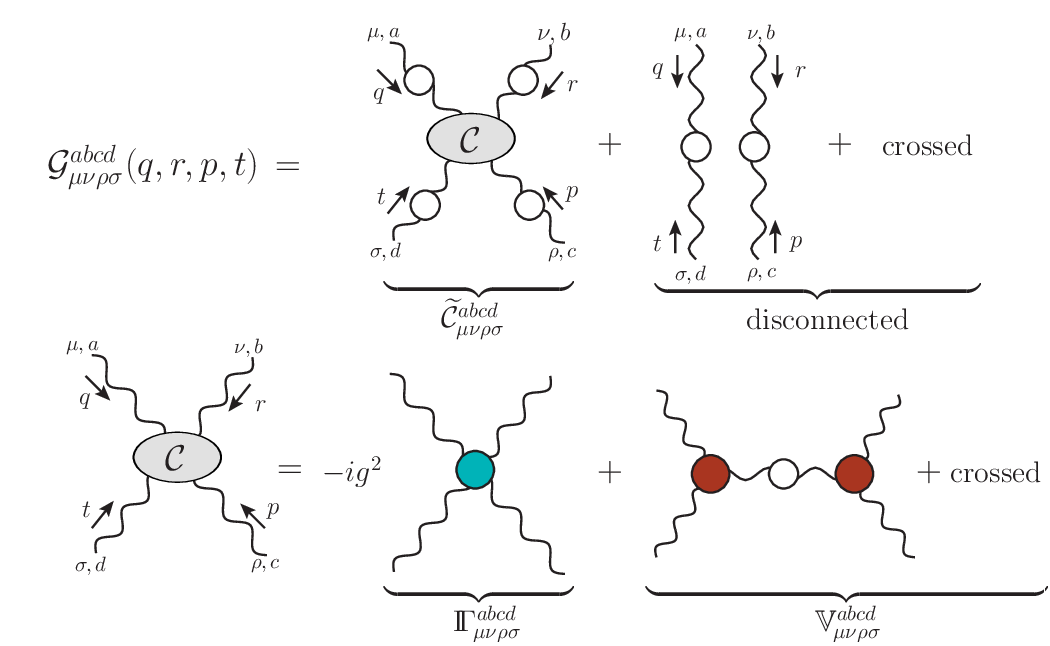}
\vspace{-0.1cm}
     \caption {Upper panel: Diagrams contributing to the full four-gluon Green's function, $\mathcal{G}^{abcd}_{\mu\nu\rho\sigma}$,  separated into connected,  $\mathcal{\widetilde C}^{abcd}_{\mu\nu\rho\sigma}$, and disconnected parts. {Lower panel:} Schematic decomposition of the \emph{amputated} four-gluon Green's function, $\mathcal{C}^{abcd}_{\mu\nu\rho\sigma}$, into the 1PI vertex, $-ig^2\fatg^{abcd}_{\mu\nu\rho\sigma}$, and the 1PR terms $\mathbb V^{abcd}_{\mu\nu\rho\sigma}$.}
     \label{fig:1P1}
\end{figure}

In general kinematics, both ${\fatg}^{abcd}_{\mu\nu\rho\sigma}$ and $\overline{\fatg}^{abcd}_{\mu\nu\rho\sigma}$ possess a multitude of
Lorentz
and color structures, leading to a large number of form factors.
In this work, we focus on the projection of the full four-gluon vertex on its tree-level structure, \ie 
\be
\Tff(q,r,p,t) :={\mathcal P}(q,r,p,t)\odot \overline\fatg (q,r,p,t) \,,
\label{eq:FormFactor}
\ee
where the symbol  ``$\odot$'' denotes the full  contraction of all Lorentz and color indices, and the projector
${\mathcal P}$ is defined as 
\be
{\mathcal P}(q,r,p,t) := \frac{ \overline{\Gamma}_{0}(q,r,p,t)}{\overline{\Gamma}_{0}(q,r,p,t)\odot\gbar_{0}(q,r,p,t)}\,.
\label{eq:projTL}
\ee
Evidently, for $\overline\fatg^{abcd}_{\mu\nu\rho\sigma} =\overline{\Gamma}_{0\,\mu\nu\rho\sigma}^{abcd}$ 
we get $\Tff=1$. 
 In addition, it is clear from \2eqs{eq:FormFactor}{eq:projTL} that $\Tff(q,r,p,t)$ is completely Bose symmetric under the exchange of any pair of its arguments. 

For the rest of this work 
we specialize to the case 
of the \emph{soft kinematics}, 
defined by setting  
$t=0$ and keeping 
the other three momenta arbitrary
but space-like, 
$q^2,r^2,p^2 <0$. 
In particular, we 
will define the 
corresponding Euclidean 
momenta $q^2,p^2,r^2 = -q^2_{\s E}, -p^2_{\s E},-r^2_{\s E}>0$, and 
drop the subscript 
``E'' throughout. 
The evaluation of the limit 
$t \to 0$
will be carried 
out
``symmetrically''~\cite{opac-b1131978NOURL},
namely
\begin{align}
    \lim_{t\to 0} \frac{t^{\sigma}t^{\sigma'}}{t^{2}} = \frac{\delta^{\sigma\sigma'}}{d} \,,  && 
    \lim_{t\to 0} P^{\sigma\sigma'}(t)  = \delta^{\sigma\sigma'}\left(1 - \frac{1}{d}\right) \,,
\label{limPS}
\end{align}
where $d$ is the dimension of space-time. 

Note that 
while the SDE determines 
directly  $\fatg^{abcd}_{\mu\nu\rho\sigma}$, the lattice computes $\mathcal{G}^{abcd}_{\mu\nu\rho\sigma}$.  
In order to extract 
$\fatg^{abcd}_{\mu\nu\rho\sigma}$ on the lattice, the redundant  contributions must be 
either eliminated 
by means of an appropriate 
choice of kinematics, or 
explicitly subtracted out. 
In particular, the disconnected contributions can be removed provided that no two momenta add up to zero, \eg $q+r\neq0$, 
and similarly for all other 
pairs of momenta;  these conditions 
eliminate propagator-like transitions, due to the non-conservation of momentum. 
As for the 1PR term, its contribution may be 
subtracted out by capitalizing on the ample  knowledge on the 
structure of the gluon propagator and 
three-gluon vertex~\cite{Bogolubsky:2009dc,Aguilar:2021okw,Eichmann:2014xya,Ferreira:2023fva,
Pinto-Gomez:2022brg,Aguilar:2023qqd,Pinto-Gomez:2024mrk}. Specifically, we 
combine 
\2eqs{eq:allbars}{eq:FormFactor}
to obtain 
\begin{align}
\Tff(q,r,p,0) := \lim_{t\to 0} \,{\mathcal P}\odot (\mathcal{\overline C} -  \overline{\mathbb V}) \,,
\label{eq:proj_noncollinear}
\end{align}
In order to subtract out 
$\overline{\mathbb V}$
we employ an excellent approximation 
for the three-gluon vertices 
contained in it. 
Specifically, we set~\cite{
Eichmann:2014xya,Ferreira:2023fva,
Pinto-Gomez:2022brg,Aguilar:2023qqd,Pinto-Gomez:2024mrk}. 
\be 
\!\!\overline{\fatg}^{\,\alpha \mu \nu}(q,r,p) =\Ls(s^2) \overline{\g}_{\!0}^{\,\alpha \mu \nu}(q,r,p) \,, \quad\!\! \!s^2= \frac{1}{2}(q^2+r^2+p^2)\,,
\label{compact}
\ee
where $\Ls(s^{2})$ is the form factor associated with the soft-gluon limit of the three-gluon vertex \mbox{$(q=0, r=-p)$}, and has been accurately determined in lattice simulations~\mbox{\cite{Athenodorou:2016oyh,Duarte:2016ieu,Boucaud:2017obn,Pinto-Gomez:2022brg,Aguilar:2019uob,Aguilar:2021lke,Aguilar:2021okw}}
and continuous studies~\mbox{\cite{Blum:2014gna,Williams:2015cvx,Aguilar:2019jsj,Aguilar:2023qqd}}. 
Implementing this approximation, we have that 
\begin{align}
\lim_{t\to 0} \,{\mathcal P}\odot \overline{\mathbb V} = f_{qrp} \, \Delta(p^2) \, \Ls(p^2)  \Ls(s^2)\, + \, \text{crossed} \,,  
    \label{eq:3gterms}
\end{align}
with 
\begin{align}
\!\!\!f_{qrp}\! =\! \frac{[5 (q^{2}+r^{2})+p^{2}]}{{72 q^{2} r^{2}}} \left[(q^{2}-r^{2}+p^{2})^2 \!-\!4q^{2}p^{2}\right]\!\,.    
\end{align}

\section{SDE analysis}
\label{SDE}

We next determine 
the form factor $\Tff(q,r,p,t)$ defined  in \1eq{eq:FormFactor} through appropriate 
projections of the 
SDE governing $\fatg_{\mu\nu\rho\sigma}^{abcd}$. In particular,   
 we employ the  formalism of the 4PI effective action~\cite{Cornwall:1974vz,Cornwall:1973ts,Berges:2004pu,York:2012ib,Williams:2015cvx} at the {\it four-loop} level~\cite{Carrington:2010qq,Carrington:2013koa}; the diagrammatic 
representation of the resulting SDE is given in \fig{fig:SDE4g}. Note that the dotted lines carrying an arrow in diagram ($d_1$)
denote the ghost propagator, 
$D(q^2) = i F(q^2)/q^2$, where $F(q^2)$
is the ghost 
dressing function, while
the 
dark-blue circles 
stand for the fully-dressed ghost-gluon vertices. 

\begin{figure}[h]
\includegraphics[width=1.0\linewidth]{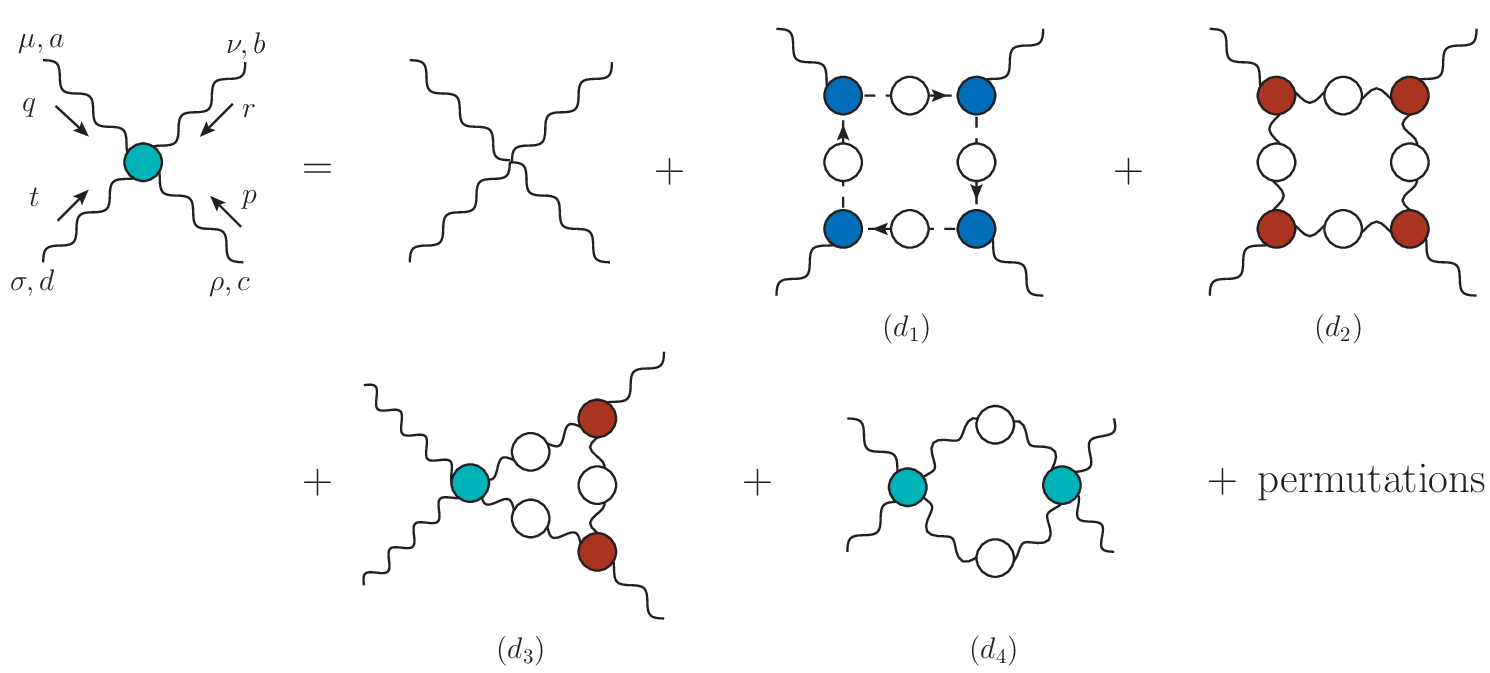}
\vspace{-0.7cm}
\caption{Diagrammatic representation of the one-loop dressed SDE for the full four-gluon vertex, ${\fatg}^{abcd}_{\mu\nu\rho\sigma}$, derived from the 
four-loop 4PI effective action. We omit contributions obtained through permutations of the external legs.}
\label{fig:SDE4g}
\end{figure}

Then, contracting both sides of the SDE  
by the projector ${\cal T}^{\mu'\nu'\rho'\sigma'}_{\mu\nu\rho\sigma}$ given 
in \1eq{projP}, we get [suppressing the arguments 
$(q,r,p,t)$]
\begin{align}
	\label{4gsde}
	\overline{\fatg}^{abcd}_{\, \mu\nu\rho\sigma} = \overline{\Gamma}_{\!0\,\mu\nu\rho\sigma}^{abcd} \,
	+  \sum_{i=1}^{4} \left(\bar{d}^{s}_{i}\right)^{abcd}_{\mu\nu\rho\sigma}\,,
\end{align}
with
\begin{align}
(\bar{d}_{i}^{s})^{abcd}_{\mu\nu\rho\sigma}:=
{\cal T}^{\mu'\nu'\rho'\sigma'}_{\mu\nu\rho\sigma} ({d}_{i})^{abcd}_{\mu'\nu'\rho'\sigma'}
+  \cdots\,,
\end{align}	
where the ellipsis denotes the 
permutations corresponding to 
each graph (not shown in 
\fig{fig:SDE4g}).

The renormalization of this SDE 
is implemented multiplicatively,  
following standard procedures.
Due to the fact that 
all vertices in the diagrams of 
\fig{fig:SDE4g} are fully-dressed, 
the only renormalization 
constant that survives is that of 
the four-gluon vertex, defined through 
$\fatg^{abcd}_{\!\!\s R\,\mu\nu\rho\sigma} =  Z_4 \fatg^{abcd}_{\mu\nu\rho\sigma}$~\cite{Aguilar:2024fen}. In particular, 
\begin{align}
	\label{renorm_4gsde}
	\overline{\fatg}^{abcd}_{{\!\! R}\, \mu\nu\rho\sigma}= Z_4\overline{\Gamma}_{\!0\,\mu\nu\rho\sigma}^{abcd} \,
	+  \sum_{i=1}^{4} \left(\bar{d}^{s}_{i\,\s R}\right)^{abcd}_{\mu\nu\rho\sigma}\,,
\end{align}
where the index ``$R$'' in  
$(\bar{d}^{s}_{i\,\s R})$
indicates that all ingredients comprising this 
set of diagrams have been replaced by their renormalized counterparts.

Note that in order to derive the expression for $\Tff(q,r,p,t)$ in the \emph{soft kinematics} defined in the introduction, one has to act  on Eq.~\eqref{renorm_4gsde} with the projector given in Eq.~\eqref{eq:projTL}, and in the sequence take the limit \mbox{$t\to 0$}, with the help of Eq.~\eqref{limPS}. After doing these steps, we find that 
\be
\Tff(q,r,p,0) = 
Z_4 +  \lim_{t\to 0}  \,\sum_{i=1}^{4} \mathcal{P}\odot \left(\bar{d}_{i}^{s}\right)(q,r,p,t)\,,
\label{D4g_eqs}
\ee
where we suppress the index “$R$” to avoid notational clutter.

 In general kinematics, 
the principal variable 
for exploring   
the degree of planar
degeneracy displayed by the 
four gluon vertex is 
\begin{align}
\sfg^{2} =  \frac{1}{2}(q^2+r^2+p^2+t^2)\,, 
	\label{planarG1}
\end{align}
while the corresponding variable for the three-gluon vertex is the $s^2$ of 
\1eq{compact}. Evidently, 
in the soft configuration 
($t=0$) the $\sfg^{2}$ of 
\1eq{planarG1}
reduces to the $s^2$ of 
\1eq{compact}. 

In this limit, in addition to 
the $s^2$, it is convenient 
to introduce two supplementary  
kinematic variables, $\x$ and $\y$, defined by~\cite{Eichmann:2014xya,Pinto-Gomez:2022brg,Aguilar:2023qqd}\footnote{The $\x$ and $\y$ are related to the $a$ and $b$ of \cite{Pinto-Gomez:2022brg} by $a = -\y$ and $b = \x$.}
\begin{align}
\x &:= \frac{\sqrt{3}(r^2 - q^2)}{2 s^2} \,, \quad 
\y := \frac{q^2 + r^2 - 2 p^2}{2 s^2} \,, 
\label{Pgroup_variables}
\end{align} 
where, due to momentum conservation, $\x$ and $\y$ are constrained to the unit disk $\x^{2} + \y^{2} \leq 1$. Additionally, it is convenient to employ polar coordinates $(R,\phi)$ defined by
\begin{align}
    R=(x^{2}+y^{2})^{\!1/2}\,, && \phi=\arctan(y/x)\,.
\end{align}
Hence, with this change of variables we have \mbox{$\Tff(q,r,p,0) \to \Tffarg$}. 

By appealing to the Bose-symmetry of the four-gluon vertex and the  
analysis of~\cite{Pinto-Gomez:2024mrk},
one may show that  
only one sixth of this disk  is 
relevant;  
the  remaining five regions can be obtained by applying to $\Tffarg$  simple transformations derived from \1eq{Pgroup_variables}.
In the numerical analysis, we isolate one such region by restricting  $\phi$ to satisfy the constraint\footnote{This constraint is equivalent to taking values of $\x$ and $\y$ satisfying the condition \mbox{$0  \leq \sqrt{3}/{3}\x\leq  \y$}.}  ${\pi}/{6} \leq \phi\leq {\pi}/{2}$. 
In particular, we have defined a grid of configurations over this region as shown in \fig{fig:disk}. The grid is defined by taking the line with $\phi=\pi/3$, and taking points over this and parallel lines so that the result is uniformly distributed.

The approximation we employ for the $\overline{\fatg}^{abcd}_{\mu\nu\rho\sigma}$ present on the rhs of Eq.~\eqref{D4g_eqs}  is analogous to the planar degeneracy relation shown in \1eq{compact} for the three-gluon vertex. 
Namely, we take 
\begin{align}
	\overline{\fatg}_{\mu\nu\rho\sigma}^{abcd}(q,r,p,t)
	& \approx 
        \Tff^{*}(\sfg^{2})
        \overline{\Gamma}_{\!0~\!\!\mu\nu\rho\sigma}^{abcd}(q,r,p,t)\,,
	\label{approx}
\end{align}
where the form of $\Tff^{*}(\sfg^{2})$ will be determined  through an iterative process.

By substituting \2eqs{approx}{planarG1} in the SDE~\eqref{renorm_4gsde}, $\Tffarg$ reads
\begin{align}
	\!\!\!\! \Tffarg = Z_{4} + \left[\int_{k}\!\! K_{1} + \!\!\int_{k}\!\! \Tff^{*} K_{2}  + \!\!\int_{k}\!\! \Tff^{*2} K_{3}\right] \,, \label{eqD}
\end{align}
for kernels $K_{i}$, whose form will not be specified here.

In order to determine the renormalization constant $Z_4$, we employ a variant of the  momentum subtraction (MOM) scheme~\cite{Boucaud:2008gn,Boucaud:2011eh},
defined through the 
condition 
\begin{align}
    \Tff(s^{2},R^{\,0},\phi^{\,0})\rvert_{s^{2}=\mu^{2}}  = 1\,,
    \label{MOM}
\end{align}
where $\mu$ is the renormalization point, 
and the set $(R^{0},\phi^{0})$ defines a particular 
kinematic configuration on the disk of 
\fig{fig:disk}. 
Applying \1eq{MOM} on \1eq{eqD}, 
we find 
\begin{align}
    Z_{4} = 1 -\left[ \int_k\!\!K_{1} + \!\!\int_k\!\! \Tff^{*} K_{2} + \!\!\int_k\!\!\Tff^{*2} K_{3}\right]_{\substack{\!\!\!\!\!s^2\to \mu^2 \\ R,\phi \to R^{\,0}\!\!, \phi^{\,0}}}\,\,\,.
    \label{eqDZ4}
\end{align}
 
In what follows 
we choose $(R^{0},\phi^{0}) = (0.71, \pi/3)$, highlighted with a black circle in the central part of \fig{fig:disk}; this point lies on the 
aforementioned grid, and is 
near the center of the region. This choice is arbitrary, and we have confirmed that using different configurations for renormalization procedure only change our results by a multiplicative factor.  
In addition, we choose
\mbox{$\mu = 4.3$~GeV}.

For the numerical evaluation of the SDE we employ the following inputs.
For the gluon propagator, $\Delta(r^2)$, and ghost dressing function, $F(r^2)$, we use the fits to the lattice results of~\cite{Bogolubsky:2009dc,Aguilar:2021okw} given by  Eqs.~(C11) and (C6) of~\cite{Aguilar:2021uwa}, respectively. 
For the transversely projected ghost-gluon vertex we use the SDE results of    
~\cite{Aguilar:2022thg,Ferreira:2023fva}, while 
for the three-gluon vertices 
we employ \1eq{compact},
with $\Ls(r^{2})$, given by a fit to the lattice data of~\cite{Aguilar:2021lke}
expressed by Eq.~(C12) in~\cite{Aguilar:2021uwa}.
Both vertices have been consistently renormalized by employing Eqs.~(B6) and (B7) of~\cite{Aguilar:2024fen}. 
Finally, we use \mbox{$\alpha_s(\mu^2) = g^2/4\pi = 0.27$}, as obtained in Sec.~\ref{subsec:eff}.

The iterative process for 
determining $\Tffarg$
may be summarized as follows:

{\it(i)} The initial 
input for the $\Tff^{*}$ on the rhs of \1eq{eqD} is 
simply its tree-level value, 
\ie $\Tff^{*} \to 1$. 

{\it(ii)}  $\Tffarg$ is then determined through the numerical integration of \2eqs{eqD}{eqDZ4}. 
 For  $s^2$ we employ a 
 grid distributed 
logarithmically over the interval $[10^{-4},10^{4}]~\text{GeV}^{2}$, whereas $R$ and $\phi$ are evaluated on the $N=491$ points of the grid sketched in \fig{fig:disk}.

%
\begin{figure}[t]
\begin{center}
\hspace{-0.9cm}\includegraphics[width=0.82\linewidth]{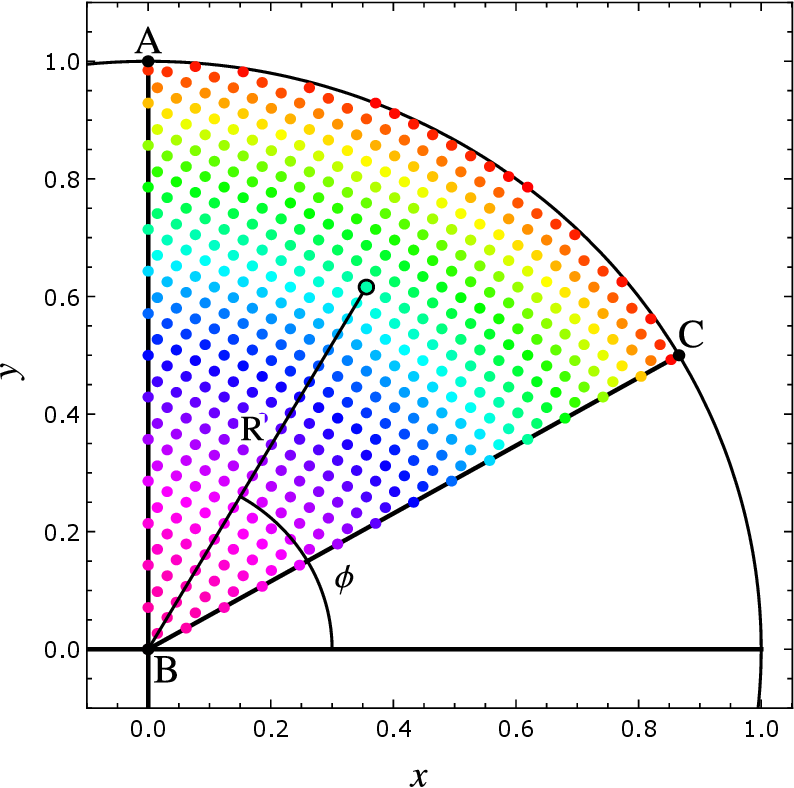} 
\end{center}
\vspace{-0.5cm}
\caption{Representation of the 491 kinematic configurations computed on 
the ($\x$, $\y$) plane. The black points $A$ and $C$ denote the collinear configurations $(p,-p,0,0)$ and $(p,p,-2p,0)$, respectively, while $B$  is the symmetric configuration defined by $p_{1}^{2}=p_{2}^{2}=p_{3}^{2}$. We also highlight in polar coordinates $(R,\phi)$ the configuration 
$(R^{0},\phi^{0})=(0.71, \pi/3)$ used for renormalizing the SDE.}
\label{fig:disk}
\end{figure}

\begin{figure}[t]
\begin{center}
\hspace{-0.5cm}\includegraphics[width=1.01\linewidth]{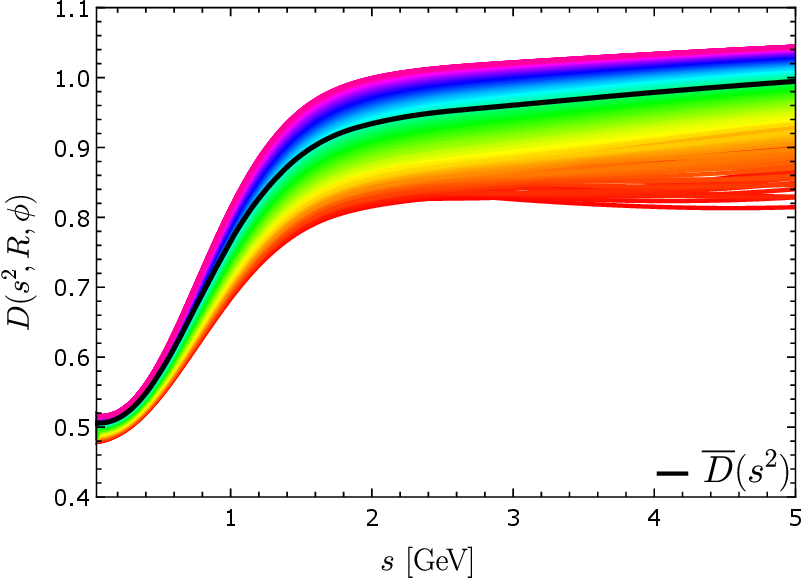} 
\end{center}
\vspace{-0.5cm}
\caption{ The form factor $\Tffargi$  for all configurations studied, and the corresponding average $\Dav(s^{2})$ (black curve). Note that the color code used corresponds 
to that of the associated configuration in the kinematic disk in~\fig{fig:disk}.}
\label{fig:conf}
\end{figure}
%

{\it(iii)} Then, we compute the simple average of these $N$ configurations by
\begin{align}
\Dav(s^{2}) = \frac{1}{N}\sum_{i=1}^{N} \Tffargi\,, 
\label{eq:av}
\end{align}
which will be used as the ``seed'' for the next iteration. Specifically, we replace $\Tff^{*}\to \Dav$ into \2eqs{eqD}{eqDZ4} and re-compute $\Tffarg$ for the same values of the external grid. 

{\it(iv)} The iterative procedure outlined above is repeated, and at each step, the average, $\Dav(s^{2})$, defined in Eq.~\eqref{eq:av}, is calculated.  Our convergence criterion is defined when the relative error between two consecutive averages is smaller than $0.1\%$. 

{\it(v)} When this is achieved,  we use the last $\Dav(s^{2})$ as input to obtain  the final results for $\Tffargi$.

The results for both $\Tffargi$ and $\Dav(s^{2})$ are shown in \fig{fig:conf}. Each configuration corresponds to a choice of $(R_{i},\phi_{i})$ and is color coded according to its position in the kinematic disk of \fig{fig:disk}, while the average is shown in black. 

Notice that there is a clear pattern between the position of a configuration on the disk and its overall behavior with respect to $\Dav(s^{2})$: configurations closer to the center are above the average, while those closer to the edge are below it.  We emphasize that the majority of the 491 curves are located very close to the total average, see Sec.~\ref{subsec:planar} for details. This observation indicates the extent to which the form factor satisfies planar degeneracy, \ie 
\begin{align}\label{softpd}
    \Tffarg\approx\Dav(s^{2})\,.
\end{align}
A detailed analysis of this relation is given in Sec.~\ref{subsec:planar}.

\section{Vertex form factors from lattice QCD}

In order to obtain
lattice results for the  four-gluon vertex in the soft kinematics,  
we  have exploited 
10.000 
quenched lattice gauge field configurations in the Landau gauge, whose set-up parameters are (the number of configurations within parenthesis): 
  $\beta=5.6$ (2.000), $5.7$ (1.000), $5.8$ (2.000), $6.0$ (2.000), $6.2$ (2.000), and $6.4$ (1.000) for $L/a=32$. The lattice spacings have been obtained using the absolute calibration~\cite{Necco:2001xg} at $\beta=5.8$, and a relative calibration based on the gluon propagator scaling~\cite{Boucaud:2018xup} for the rest of the $\beta$'s.
For each set of momenta, we have exploited the full discrete lattice symmetry to average over all equivalent momenta with respect to permutations of Lorentz indices or signs among components. Moreover, as the lattice artifacts that break rotational symmetry (termed H4-errors) are typically far smaller than the statistical errors associated with three- or four-point correlation functions, we will average together all the sets of momenta that differ by higher-order H4 
invariants~\cite{deSoto:2007ht,deSoto:2022scb,Pinto-Gomez:2024mrk}. 

Following the 
analysis of Sec.\ref{sec:tensorial},  we use 
\2eqs{eq:proj_noncollinear}{eq:3gterms} to remove from $\Tff(q,r,p,0)$ the 
1PR contributions. The 
required subtraction is carried out 
{\it individually} for each kinematic configuration, 
using the available lattice data for the 
functions 
$\Delta(r^2)$ 
and $\Ls(r^2)$. 
Subsequently, the unrenormalized data sharing the same value of the Bose invariant $s^2$ are averaged, defining the quantity $\DavL(s^{2})$.

The renormalization procedure 
employed on the lattice 
is different from that used in the 
SDE analysis, \ie  
the MOM condition \1eq{MOM}; however, the 
the two schemes coincide in the limit 
of exact planar degeneracy.
Specifically, on the lattice the  multiplicative renormalization constant,  $Z_4$ [defined right before \1eq{renorm_4gsde}] 
is fixed by imposing on the averaged data
 the condition
 \begin{align}
\DavL^{\s R}(s^{2}) &=  Z_4 \DavL(s^{2})\,, \quad \DavL^{\s R}(\mu^{2}) = 1\,, 
\label{Lat-avg}
\end{align}
with the renormalization point $\mu = 4.3~\mbox{GeV}$. 
As was done with the SDE results, 
in what follows
we  suppress 
the suffix ``$R$''.

Note that all lattice 
errors are statistical, computed through the application of the  ``Jack-knife method''. Moreover,  
the systematic errors stemming from 
the assumption of perfect 
planar degeneracy are  
subleading compared to the statistical; 
this is corroborated by 
the smooth behavior and small dispersion exhibited by the data for $\DavL$, displayed in terms of $s^2$ in Fig.\,\ref{fig:lattice}.
The same is true for the 
errors associated with the 
continuum limit; 
therefore,  
the dependence on the lattice spacing $a$ has been 
suppressed in 
\1eq{Lat-avg}.

In order to perform a meaningful comparison with the SDE-derived average of \fig{fig:conf}, the form factor $\Dav(s^{2})$ is rescaled in order to match the lattice renormalization scheme of \1eq{Lat-avg}. This is accomplished through 
the operation \mbox{$\Dav(s^{2})\to\Davn(s^{2}):=\Dav(s^{2})/\Dav(\mu^{2})$}, where $\Davn(s^{2})$ denotes the \emph{normalized average}, satisfying $\Davn(\mu^{2}) = 1$.

In addition, we introduce a band surrounding the SDE-derived results indicating uncertainties associated with the tree-gluon form factor $\Ls(r^2)$. This is implemented by repeating the iterative procedure outlined in the previous section, and solving Eq.~\eqref{eqD} numerically using as input for $\Ls(r^2)$ the band defined by Eq.~(C13)  in~\cite{Aguilar:2021uwa}.

\begin{figure}[t]
  \includegraphics[width=0.95\linewidth]{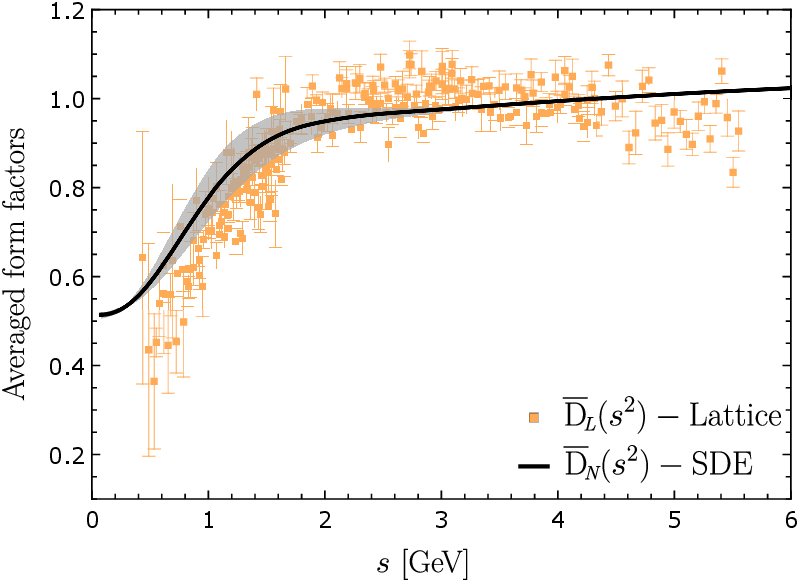} 
  \vspace{-0.3cm}
\caption{ Comparison between the averaged form factor obtained from the lattice, $\DavL(s^{2})$, and the normalized average from the SDE, $\Davn(s^{2})$. The band around $\Davn(s^{2})$ denotes the propagation of the statistical error associated with the three-gluon vertex.}
\label{fig:lattice}
\end{figure}

In 
Fig.~\ref{fig:lattice}, 
the 
lattice results for $\DavL(s^{2})$ are shown 
for all values of $\beta$, and are 
compared to 
$\Davn(s^{2})$; we have a total of $N_{\srm{lat}} = 250$ points. We 
observe that the averaged form factors computed with both methods are quantitatively rather similar. The discrepancy between both results may be measured by the mean absolute percentage error, $\sigma$, \ie
\begin{align}
     \sigma = \frac{1}{n} \sum_{i=1}^n \left|\frac{
    \Davn(s_{i}^{2}) - \DavL(s_{i}^{2})}{\DavL(s_{i}^{2})}\right| \!\times\!100\%\,,
    \label{eq:sigma}
\end{align}
with $s_{i},\,i\!=\!1,\ldots,n$, denoting the momenta $s$ of the $n$ lattice data points within a certain interval in \fig{fig:lattice}. Considering the complete data set, with $n = N_{\srm{lat}}$, we find $\sigma=7\%$. As is evident from \fig{fig:lattice}, the largest deviations between $\Davn(s_{i}^{2})$ and $\DavL(s_{i}^{2})$ occur for \mbox{$s < 1~\text{GeV}$}, for which we find (with $n= 29$) $\sigma=19\%$, and \mbox{$s > 4.5~\text{GeV}$}
(with $n= 20$),  
which yields $\sigma=8\%$.

Both results display an infrared suppression with respect to the tree-level value (unity),  remaining positive for the entire range of momentum. 
Note that, at 
the origin, the SDE result reaches a finite value; in fact, one may check the finiteness by directly setting $s=0$ in \1eq{eqD}, and taking the limit when all momenta vanish. Instead, the lattice
appears inconclusive on this matter, because the data in that region 
come with large errors, and 
do not reach below $s\leq 0.5~\text{GeV}$.
In addition, one can also see that the expected error in $\Ls(r^2)$ has only a mild effect on $\Davn(s^{2})$ in the region $s\in [0.4, 3]~\text{GeV}$.

Finally, we note that SDE and lattice select their
kinematic configurations differently: in contrast to the random sampling obtained in the Monte Carlo method, \fig{fig:disk} is composed of uniformly spaced points.  However, given the relatively high number of configurations probed, together with the smoothness of $\Tffarg$ observed in \fig{fig:conf}, the discrepancy introduced by this difference is minimal.

\section{Deviations from planar degeneracy}
\label{subsec:planar}
In this section we investigate the accuracy of the planar degeneracy, as expressed 
through \1eq{softpd}, and propose an adjustment 
that provides a more accurate approximation for this form factor.

To accomplish this, we compute the relative percentage deviation of $\Tffargi$ from the total average, $\Dav(s^{2})$, through the relative deviation
\begin{align}
    \delta^{i}(s^{2}) \!=\! \left|\frac{\Tffargi -\Dav(s^{2})}{\Tffargi}\right|\!\times\!100\%, \quad i\!=\!1,\cdots\!,491\,.
    \label{eq:error-s}
\end{align}

For \mbox{$s\leq 1~\text{GeV}$}, 
we find that the maximum error \mbox{$\delta^{i}(s^{2})\leq 12\%$}. It is clear from~\fig{fig:conf} 
that as $s$ grows, 
the separation between 
curves increases; this 
tendency is captured by the  
$\delta^{i}(s^{2})$, 
which also increases 
within the range \mbox{$1 < s \leq 6~\text{GeV}$}, 
reaching a maximum value of $22\%$. 
Note, however, that for the vast majority of configurations, 
$\delta^{i}(s^{2})$ is considerably smaller:  out of the $491$ configurations analyzed, $439$ deviate from the average by less than $10\%$, while $285$  
by less than $5\%$.

%
\begin{figure}[t]
  \includegraphics[width=1\linewidth]{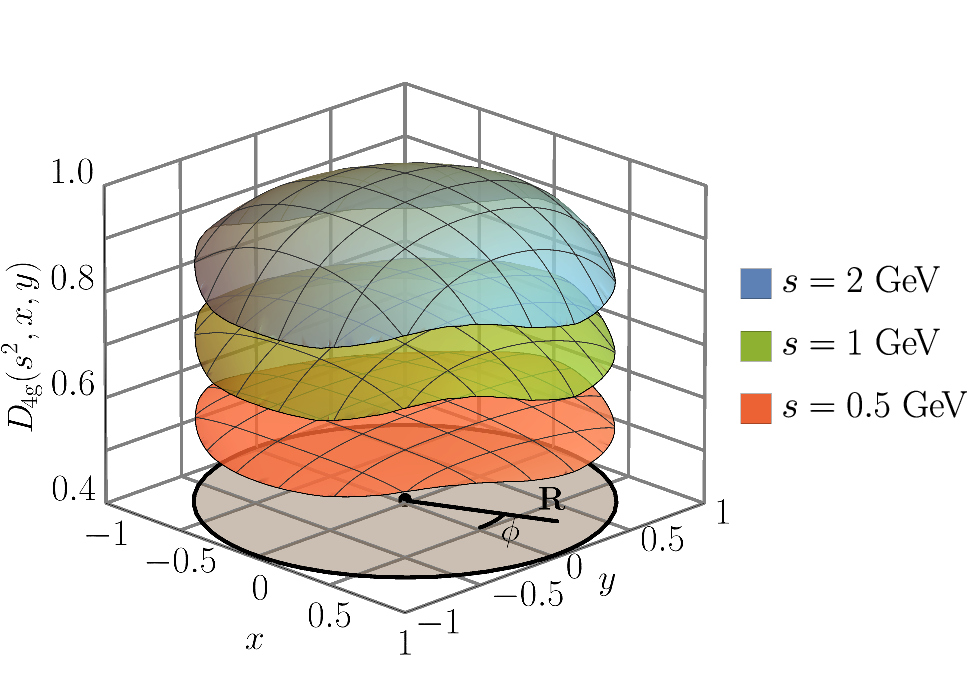}
\vspace{-0.5cm}
\caption{$\Tffarg$ plotted as a function of $\x$ and $\y$ for selected values of $s$. The  dependence on $\x$ and $\y$ is manifested by a deviation from perfect flatness, indicating a violation of planar degeneracy.}
\label{fig:pancakes}
\end{figure}

The property of planar degeneracy can also be appreciated directly by plotting $\Tffarg$ as a function of $\x$ and $\y$, and analyzing the ``flatness'' of these surfaces for a fixed value of $s^{2}$.  The result is shown in~\fig{fig:pancakes}. There, one clearly sees that for \mbox{$s = 0.5$~GeV} and  \mbox{$s = 1$~GeV}, the $\Tffarg$ is almost perfectly constant, \ie nearly independent of the variables $\x$ and $\y$.

For \mbox{$s= 2$~GeV}, one sees a bending more pronounced at the edges of the disk represented in Fig.~\ref{fig:disk}, 
and the appearance of a plateau in the internal region of the disk \ie $R\lessapprox 0.5$.

An interesting feature regarding \fig{fig:pancakes} is that the deviation from planar degeneracy, manifested in the curvature of the surfaces, 
depends almost exclusively on the radius $R$, showing  
minimal dependence 
on the angle $\phi$.

This property suggests that instead of simply using $\Dav(s^{2})$, a more precise approximation to $\Tffarg$ may be achieved by taking in consideration both the dependence on $s^{2}$ and the radius $R$.\footnote{The relevance of $R$ can be traced 
back to~\cite{Eichmann:2015nra},  since it is proportional to the 
singlet of the group $S_4$
with the 
second-lowest dimension.}

We explore this possibility by performing an average over the angle $\phi$. In particular, we interpolate the data points so that $\phi$ varies continuously in the interval $[\pi/6,\pi/2]$, and sample \mbox{$N_{\phi}=10$} equally spaced angles $\phi^{i}$, \mbox{$i =1,\ldots,N_{\phi}$}. Thus, in analogy to \1eq{eq:av}, we define
\begin{align}
    \Dav(s^{2},R) = \frac{1}{N_{\phi}}\sum_{i}^{N_{\phi}} \Tff(s^{2},R,\phi^{i})\,.
    \label{eq:Ravg}
\end{align}
Then, 
the improvement 
achieved when using \1eq{eq:Ravg} can be appreciated  by defining the relative deviation
\begin{align}
    \!\!\!\!\overline{\delta}^{\,i}(s^{2}) \!=\! \left|\frac{\Tffargi \!-\!\Dav(s^{2},R_{i})}{\Tffargi}\right|\!\times\!100\%, ~~ i\!=\!1,\cdots\!,491\,.
    \label{eq:error-sr}
\end{align}
In contrast to $\delta^{i}(s^{2})$, the relative error $\overline{\delta}^{\,i}(s^{2})$ has a maximum deviation of only $7\%$ over the entire  range \mbox{$0 < s \leq 6~\text{GeV}$}. 
Indeed, when compared to $\Tff$, with its complete momentum dependence taken into account, $\Dav(s^{2},R)$ offers a significant increase in accuracy over $\Dav(s^{2})$.

\section{Effective charge}
\label{subsec:eff}

In this section we 
use the SDE and lattice results of \fig{fig:lattice} to 
construct an RGI
effective charge, denoted by  
$\alpha_{4g}(s^{2})$. 
Specifically, following a standard definition~\cite{Kellermann:2008iw,Cyrol:2014kca,Huber:2020keu,Aguilar:2024fen}, we have 
\begin{align}\label{eq:alpha4g}
    \alpha_{4g}(s^{2}) = \alpha_{s}(\mu^{2}) \Dav(s^{2}) \mathcal{Z}^2(s^{2}) \,,
\end{align}
where $\Dav(s^{2})$ stands for the averaged form factors shown in \fig{fig:lattice}, and  
$\mathcal{Z}(p^{2}) = p^{2}\Delta(p^{2})$ 
is the gluon dressing function.

Capitalizing on the 
RG invariance 
of $\alpha_{4g}(s^{2})$, the SDE approach employs the  average of \fig{fig:conf} without the normalization factor introduced in the previous section. The value $\alpha_{s}(\mu^{2}) = 0.27$ quoted in Sec.~\ref{SDE} is obtained by minimizing the discrepancy of the SDE-lattice comparison in the vicinity of the renormalization point $\mu=4.3~\text{GeV}$; specifically this value produces the best match in the interval $s\in[4,4.5]~\text{GeV}$.

The lattice determination of $ \alpha_{4g}(s^{2})$ follows the same procedure 
described in~\cite{Pinto-Gomez:2024mrk}, except that in \1eq{eq:alpha4g} the implicit continuum limit $a\to 0$ has again been dropped because any non-singular, remaining dependence on the lattice spacing is hidden in the statistical noise.

Both determinations of  $\alpha_{4g}(s^{2})$ are shown in \fig{fig:Charge}, displaying an excellent agreement over the entire range of momenta: 
the mean absolute percentage error, defined in analogy to \1eq{eq:sigma}, is of $9\%$ for the entire interval. %
Let us finally stress 
the qualitative similarities with the result of~\cite{Aguilar:2024fen}, where the four-gluon 
form factor in \1eq{eq:alpha4g} is
evaluated in a collinear kinematic configuration.

\begin{figure}[t]
  \includegraphics[width=0.95\linewidth]{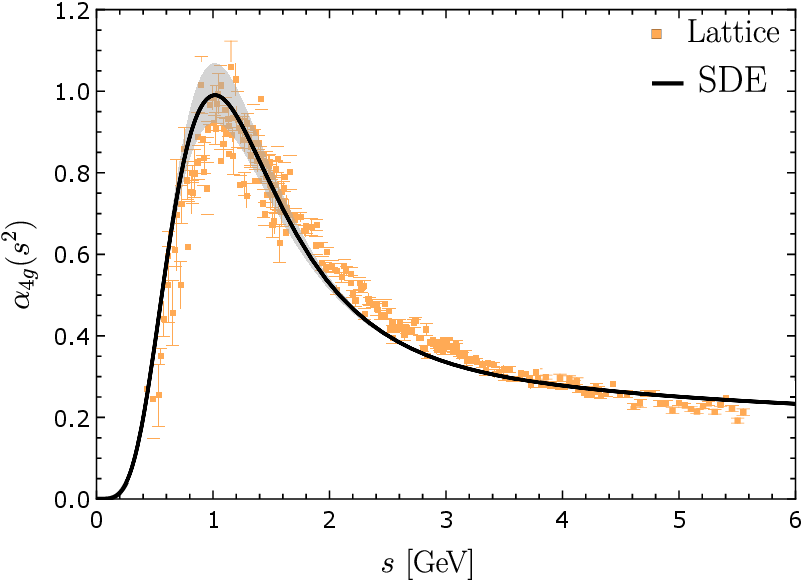} 
  \vspace{-0.3cm}
 \caption{Lattice and SDE determinations of the $\alpha_{4g}(s^{2})$ defined by \1eq{eq:alpha4g}, with error bands associated to the statistical error of the three-gluon vertex.}
\label{fig:Charge}
\end{figure}

\section{Conclusions}
In this work we have explored 
the nonperturbative four-gluon vertex in soft kinematics, 
by combining an  
SDE analysis and a lattice simulation, in the Landau gauge.
The quantity considered is 
the projection of the 
four-gluon vertex
on its tree-level tensor, 
averaged over 
a large selection of kinematic configurations sharing the same $s^{2}$.

The key hypothesis employed by both 
methods is the property of planar 
degeneracy: its use simplifies the SDE analysis, 
and allows for the emergence of a clear lattice 
signal.
 The results of both 
approaches are in very good agreement, affirming the overall 
robustness of the 
underlying picture. 

Importantly, a further analysis of the SDE results establishes that planar degeneracy is only approximate, as already argued in~\cite{Aguilar:2024fen}. Here, we have indicated how deviations of this property can be accurately taken into account, suggesting an improved description for this vertex in future applications. 

We emphasize that, 
while the quantity $\Tff(q,r,p,t)$
considered in this work is expected to 
be dominated by the classical form factor, in a future analysis  this assumption 
may be explicitly tested, by formally 
eliminating unwanted admixtures through suitable projections, in the spirit of~\cite{Aguilar:2024fen}. 

We finally point out that, in the limit of 
exact planar degeneracy, the effective charge 
$\alpha_{4g}(s^{2})$ 
would measure, in 
a configuration-independent way,  
the strength of the four-gluon
interaction.
Even though the observed deviations 
from the planar degeneracy invalidate this possibility, their reduced size  
makes $\alpha_{4g}(s^{2})$ 
a rather useful instrument  
for describing the underlying dynamics.

\vspace{-0.3cm}

\section*{Acknowledgements}
The work of  A.~C.~A. and L.~R.~S. are supported by the CNPq grants \mbox{310763/2023-1} and \mbox{162264/2022-4}, and is part of
the project INCT-FNA 464898/2014-5. 
This study was financed in part by the Coordenação de Aperfeiçoamento de Pessoal de Nível Superior - Brasil (CAPES) - Finance Code 001 (L.~R.~S.). M.~N.~F. acknowledges financial support from the National Natural Science Foundation of China (grant no. 
12135007). The research of 
J.~P. is supported by the Spanish MICINN grant PID2020-113334GB-I00,
the Generalitat Valenciana grant CIPROM/2022/66, and in part by the EMMI visiting grant of 
the ExtreMe Matter Institute EMMI
at the GSI,
Darmstadt, Germany.
F.~D.~S., J.~R.~Q. and F.~P.~G. acknowledge ﬁnancial support from the Spanish MICINN research grant PID2022-140440NB-C22. F.~P.~G. also acknowledges support from Banco Santander Universidades. The authors acknowledge the C3UPO of the Pablo de Olavide University for the support with HPC facilities.

\bibliographystyle{model1a-num-names}

\end{document}